\documentclass[12pt]{article}

\usepackage{amsmath}

\usepackage[dvipdfmx]{graphicx}

\usepackage{subfig}

\usepackage{array}

\usepackage{amsfonts}

\usepackage{epsf}

\usepackage{epsfig}

\usepackage{amssymb}

\usepackage{bbm}

\usepackage{delarray}

\usepackage{caption}

\usepackage{amstext}

\usepackage{graphics}

\usepackage{epstopdf}

\usepackage[titletoc,toc,title]{appendix}

\catcode`ð=\active
 \defð{\u{g}}
 \catcode`Ð=\active
 \defÐ{\u{G}}
 \catcode`Ý=\active
 \defÝ{\. I}
 \catcode`ö=\active
 \defö{\"{o}}
 \catcode`Ö=\active
 \defÖ{\"O}
 \catcode`ü=\active
 \defü{\"{u}}
 \catcode`Ü=\active
 \defÜ{\"{U}}
 \catcode`Þ=\active
 \defÞ{\c{S}}
 \catcode`þ=\active
 \defþ{\c{s}}
 \catcode`ý=\active
 \defý{{\i}}
 \catcode`ç=\active
 \defç{\d{c}}
 \catcode`Ç=\active
 \defÇ{\d{C}}

\numberwithin{equation}{section}
\begin{document}
\title{Klein-Gordon and Dirac Equations with Thermodynamic Quantities}

\author{{\small Altu\u{g} Arda$^a$\footnote{Present adress: Department of Mathematical Science, City University London, UK}\,\, Cevdet Tezcan$^b$\,\,
and Ramazan Sever$^c$\footnote{arda@hacettepe.edu.tr, ctezcan@baskent.edu.tr, sever@metu.edu.tr}}\\
$^a${\small \emph{Department of Physics Education, Hacettepe University}}, \\
{\small \emph{06800,
Ankara, Turkey}}\\
$^b${\small \emph{Faculty of Engineering, Baskent University, Baglica Campus, Ankara,Turkey}}
\\$^c${\small \emph{Physics Department, Middle East Technical
University}}\\ {\small \emph{06531, Ankara, Turkey}} \\}
\date{}

\maketitle

\begin{abstract}
We study the thermodynamic quantities such as the Helmholtz free energy, the mean energy and the specific heat for both the Klein-Gordon, and Dirac equations. Our analyze includes two main subsections: ($i$) statistical functions for the Klein-Gordon equation with a linear potential having Lorentz vector, and Lorentz scalar parts ($ii$) thermodynamic functions for the Dirac equation with a Lorentz scalar, inverse-linear potential by assuming that the scalar potential field is strong ($A \gg 1$). We restrict ourselves to the case where only the positive part of the spectrum gives a contribution to the sum in partition function. We give the analytical results for high temperatures.

  %\narrowtext

\end{abstract}
\newpage

\section{Introduction}
The study of the thermodynamic functions for different potential fields within
non-relativistic and/or relativistic regimes has been received a special interest for a few decades. Pacheco and co-workers have analyzed the one-dimensional Dirac-oscillator in a thermal bath [1], and then extended the same subject to three dimensional case [2]. In Refs. [3, 4], the Dirac/Klein-Gordon oscillators have been analyzed in thermodynamics point of view, and the Dirac equation on graphene has been solved to study the thermal functions, respectively. In Refs. [5, 6], the non-commutative effects on thermodynamics quantities have been discussed in graphene. The spin-one DKP oscillator has been analyzed for the statistical functions by taking into account the non-commutative effects with an external magnetic field [7]. In Ref. [8], the authors have studied the thermodynamics properties of a harmonic oscillator plus an inverse square potential within the non-relativistic region.

In the present work, we analyze some thermodynamic quantities such as the free energy, the mean energy, and the specific heat for the relativistic wave equations, namely Klein-Gordon (KG), and Dirac equations. Our basic function will be the partition function, $Z(\beta)$, of whole system. We then compute other thermal quantities, and also give the results for the case of high temperatures ($\beta \ll 1$). For the case where we deal with the Klein-Gordon equation, we obtain the analytical expressions for thermal functions by using a method based on the Euler-MacLaurin formula which has been used to obtain thermodynamics quantities of Dirac-oscillator [1, 2, 5]. Some authors, on the other hand, have used another method based on the Hurwitz zeta function to analyse the thermodynamics properties of physical systems [3, 4, 6, 9]. For our aim, we find the analytical bound state solutions of the KG equation for the linear potential with vector and scalar couplings, which is well studied in Ref. [10]. We give our main results in terms of a dimensionless parameter $\bar{\mu}$ which is written in terms of a quantity analogous to the Debye temperature in solid state physics [2]. In the second part of the present paper, we analyze the same quantities for the Dirac equation with a Lorentz scalar, inversely linear potential. We compute the partition function of the system by using a method based on the Euler and the Riemann Zeta functions, and their properties [9]. This approach has been used to handle thermal functions for different physical systems [3, 4, 11]. We consider the case where the scalar potential coupling is a large constant for getting an analytical expression for the partition function, and analyzing also the quantity numerically. As in the first part, we write all of thermal functions in terms of a dimensionless parameter, $\bar{\mu}$. Throughout the paper, we consider only the case where the particle-particle interactions are excluded for handling the partition function which means that the positive-energy levels give a contribution only, and the partition function does not involve a sum over negative-energy levels [12].

The paper is organized as follows. In Section II, we first compute the analytical solutions of the KG equation for a one-dimensional, linear potential mixing of vector and scalar part with unequal magnitudes. After determining the partition function for the case where only scalar part of the potential gives a contribution, we give some thermodynamics functions such as the Helmholtz free energy ($F(\beta)$), the mean energy ($U(\beta)$), the entropy ($S(\beta)$), and the specific heat ($C(\beta)$) with figures showing the variation of these functions versus temperature. In Section III, we solve the Dirac equation for an inverse linear potential, and derive the analytical results which will be used to write the partition function. We obtain the partition function with the help of the Euler and the Riemann Zeta functions, and then the other thermal functions. We find all thermodynamic quantities in terms of the coupling constant. So, we summarize our results in figures showing the variation of the above thermal functions versus the temperature for different values of $A$. In last Section, we give the conclusion.

\section{T\lowercase{he} K\lowercase{lein}-G\lowercase{ordon} E\lowercase{quation}}

\subsection{B\lowercase{ound} S\lowercase{tates}}

The Klein-Gordon equation for a particle with mass $m$ in the presence of a vector, $V_{1}(x)$, and a scalar potential, $V_{2}(x)$, is written as [13]
\begin{eqnarray}
\frac{d^2\psi(x)}{dx^2}+Q^2\left[(E-V_{2}(x))^2-(\mu-V_{1}(x))^2\right]\psi(x)=0\,,
\end{eqnarray}
with $Q^2=1/\hbar^2c^2$, and $\mu=mc^2$. By taking the vector and scalar part of the linear potential as $\hbar ca_{2}|x|$, and $\hbar ca_{1}|x|$, respectively, and inserting into Eq. (2.1), we can write Eq. (2.1) as a Schrodinger-like equation
\begin{eqnarray}
\left[-\frac{1}{2\mu}\frac{d^2}{dx^2}+Q^2\left(\frac{1}{2}k_{1}x^2+k_{2}|x|+\epsilon\right)\right]\psi(x)=0\,,
\end{eqnarray}
where
\begin{eqnarray}
k_{1}=\frac{a^2_{1}}{\mu Q^2}(1-a^2)\,\,; k_{2}=\frac{a_{1}}{\mu\sqrt{Q\,}}(Ea-\mu)\,\,;\epsilon=\frac{\mu^2-E^2}{2\mu}\,,
\end{eqnarray}
with $a=a_{2}/a_{1}$. Throughout the first part of the paper, we consider only the case where $a^2_{1} \neq a^2_{2}$, otherwise we have the Coulomb problem which will not be the subject of the present work. The reader can find a detailed analyze about the Coulomb case in Ref. [10]. Defining a new variable $y=|x|+k_{2}/k_{1}$ in Eq. (2.2) gives us
\begin{eqnarray}
\left[-\frac{1}{2\mu}\frac{d^2}{dy^2}+Q^2(\frac{1}{2}k_{1}y^2-\epsilon_{1})\right]\psi(y)=0\,,
\end{eqnarray}
where $\epsilon_{1}=-\epsilon+k^2_{2}/2k_{1}$. Eq. (4) corresponds to the Schrodinger equation for the harmonic oscillator with the energy $\hbar\omega(n+1/2)$ which together with Eq. (2.3) gives the bound state solutions of the linear potential as
\begin{eqnarray}
E=a\mu \pm \sqrt{(1-a^2)\mu\hbar\omega(2n+1)\,}\,,
\end{eqnarray}
Here, if one takes the frequency of the system as $\omega=\sqrt{k_{1}c^2/\mu\,}$, then our result in Eq. (2.5) is consistent with the ones obtained in literature [10].

In next section, we compute some thermal quantities of the system starting from the partition function while we deal with only particles corresponding to the positive energy levels to write the thermodynamics quantities.

\subsection{T\lowercase{hermal} Q\lowercase{uantities}}
The partition function of the system is given as [1]
\begin{eqnarray}
Z(\beta)=\sum_{n=0}^{\infty}e^{-\beta(E_{n}-E_{0})}=e^{\mu\beta}\sum_{n=0}^{\infty}e^{-\beta E_{n}}\,,
\end{eqnarray}
where $\beta=1/k_{B}T$, $k_{B}$ Boltzmann constant, and $T$ is temperature in Kelvin. With the help of Eq. (2.5), we study the following thermal quantities such as the Helmholtz free energy, the mean energy, the entropy, and the specific heat defined in terms of the partition function as following
\begin{eqnarray}
F(\beta)=-\frac{1}{\beta}\,\text{ln}\,Z(\beta); U(\beta)=-\frac{\partial}{\partial\beta}\,\text{ln}\,Z(\beta); S(\beta)=k_{B}\beta^2\frac{\partial}{\partial\beta}F(\beta); C(\beta)=-k_{B}\beta^2\frac{\partial}{\partial\beta}U(\beta)\,,\nonumber\\
\end{eqnarray}

Although the summation over $n$ in Eq. (2.6) is infinite, one can check the divergence of the summation by using the integral formula [1]
\begin{eqnarray}
\int_{0}^{\infty}e^{-\beta\sqrt{\beta'n+\beta''\,}}dn=\frac{2}{\beta'\beta^{2}}(1+\beta\sqrt{\beta''\,})e^{-\beta\sqrt{\beta''\,}}\,,
\end{eqnarray}
which means that the partition function is convergent. Because Eq. (2.6) is convergent, the partition function can be computed by using the Euler-MacLaurin formula [1, 2, 5]
\begin{eqnarray}
\sum_{m=0}^{\infty}f(m)=\frac{1}{2}f(0)+\int_{0}^{\infty}f(x)dx-\sum_{i=1}^{\infty}\frac{1}{(2i)!}B_{2i}f^{(2i-1)}(0)\,,
\end{eqnarray}
where $B_{2i}$ are the Benoulli numbers, $B_{2}=1/6$, $B_{4}=-1/30$, $\ldots$. Up to $i=2$, Eqs. (2.9) and (2.10) gives the partition function of the system as
\begin{eqnarray}
Z(\bar{\mu})=\frac{1}{2}+\bar{\mu}(\bar{\mu}+1)+\frac{1}{240\bar{\mu}^{3}}(19\bar{\mu}^{2}-\bar{\mu}-1)\,,
\end{eqnarray}

With the help of the above result, we write explicitly the thermal functions studying here in terms of a dimensionless parameter $\bar{\mu}=1/\mu\beta=k_{B}T/Mc^2$
\begin{eqnarray}
\bar{F}&=&\frac{F}{\mu}=-\bar{\mu}\text{ln}Z(\bar{\mu})\,,\nonumber\\
\bar{U}&=&\frac{U}{\mu}
=\frac{3\bar{\mu}(1+2\bar{\mu}-19\bar{\mu}^{2}+240\bar{\mu}^{4}+480\bar{\mu}^{5})}
{-1-3\bar{\mu}+57\bar{\mu}^{2}+360\bar{\mu}^{3}+720\bar{\mu}^{4}(1+\bar{\mu})}\,,\nonumber\\
\bar{S}&=&\frac{S}{k_{B}}=\frac{3(1+2\bar{\mu}-19\bar{\mu}^{2}+240\bar{\mu}^{4}+480\bar{\mu}^{5})}
{-1-3\bar{\mu}+57\bar{\mu}^{2}+360\bar{\mu}^{3}+720\bar{\mu}^{4}(1+\bar{\mu})}+\text{ln}Z(\bar{\mu})\,,\nonumber\\
\bar{C}&=&\frac{C}{k_{B}}=\frac{1}{[-1-3\bar{\mu}+57\bar{\mu}^{2}+360\bar{\mu}^{3}+720\bar{\mu}^{4}(1+\bar{\mu})]^{2}}\nonumber\\ &\times&
\left\{3(-1-4\bar{\mu}-6\bar{\mu}^{2}-606\bar{\mu}^{3}-5163\bar{\mu}^{4}-11520\bar{\mu}^{5}+43200\bar{\mu}^{6}
+309600\bar{\mu}^{7}\right.\nonumber\\&&\left.+691200\bar{\mu}^{8}(1+\bar{\mu})+345600\bar{\mu}^{10})\right\}\,,
\end{eqnarray}

Let us first analyze the case of high temperatures corresponding to $\beta \ll 1$. Eq. (2.11) gives the following results for high temperatures
\begin{eqnarray}
Z \sim \bar{\mu}^{2}\,\,;U \sim 2\bar{\mu}\,\,;C \sim 2\,
\end{eqnarray}
Here, we can stress that the behaviour of the partition function, the mean energy, and the specific heat are identical to those obtained for the relativistic oscillators [3-5]. It is basically related with the nature of the interactions both of Dirac-KG oscillators, and the linear potential considered here, namely, all of these interactions are linear in spatially coordinate.

We give our basic results in Figs. 1-5 where we plot the variation of all thermal functions according to the dimensionless parameter $\bar{\mu}$ for $Mc^2/\hbar\omega=1$. This relation between energy and frequency of the system corresponds to the relativistic case [3] where the thermal functions of the Dirac and KG oscillators have been analysed by taking into account the different values of this ratio. In Figs. 1, 3, and 4, we can see that $\text{ln}Z(\bar{\mu})$, the mean energy, and the entropy increase with temperature. In Fig. 2, we observe that the free energy decreases with temperature as expected. At this point, we should point out that the entropy of the system has a 'turning point' about $\bar{\mu} \sim 0.5$, the values of $\bar{\mu}$ smaller than $\sim 0.5$ means the non-relativistic region. This situation has been well discussed in Ref. [3], and our result for the entropy is consistent with the ones obtained in Ref. [3]. In Fig. 5, we see that he specific heat has an upper limit for high temperatures as we obtained above. We have to mention that all of these results are identical to those obtained for relativistic Dirac and KG oscillators because the interactions in Dirac (KG) oscillator(s) and here have the similar behaviour.

\section{T\lowercase{he} D\lowercase{irac} E\lowercase{quation}}

\subsection{B\lowercase{ound} S\lowercase{tates}}
The Dirac equation is written in terms of the time-independent Lorentz scalar, $V_{1}(x)$, and Lorentz vector, $V_{2}(x)$, potentials as [14]
\begin{eqnarray}
\left[c\alpha\hat{p}+\beta(mc^2+V_{1}(x))+V_{2}(x)\right]\psi(x)=E\psi(x)\,,
\end{eqnarray}
where $E$ is the energy of the fermion with mass $m$, $c$ is the speed of light, and $\hat{p}$ is the momentum operator. $\alpha$ and $\beta$ are traceless matrices satisfying the relations $\alpha^2=\beta^2=1$, and $\{\alpha,\beta\}=0$, separately. By choosing them as Pauli matrices as $\alpha=\sigma_{2}$, and $\beta=\sigma_{1}$ [15] and taking the spinor with upper and lower components as
\begin{eqnarray}
\psi(x)=\begin{pmatrix}
f(x)\\
g(x)
\end{pmatrix}\,,
\end{eqnarray}
Eq. (3.1) gives two coupled equations in the absence of the vector potential
\begin{eqnarray}
(mc^2+V_{1}(x))g(x)-\hbar c\frac{dg(x)}{dx}=Ef(x)\,,\nonumber\\
(mc^2+V_{1}(x))f(x)+\hbar c\frac{df(x)}{dx}=Eg(x)\,,
\end{eqnarray}
We obtain two second order differential equations as
\begin{eqnarray}
\left\{-\frac{d^2}{dx^2}+Q^2(\mu^2-E^2)+2Q^2\mu V_{1}(x)+Q^2V^{2}_{1}(x)-Q\frac{dV_{1}(x)}{dx}\right\}f(x)=0\,,\nonumber\\
\left\{-\frac{d^2}{dx^2}+Q^2(\mu^2-E^2)+2Q^2\mu V_{1}(x)+Q^2V^{2}_{1}(x)+Q\frac{dV_{1}(x)}{dx}\right\}g(x)=0\,,
\end{eqnarray}
with $Q=1/\hbar c$, and $\mu=mc^2$. Inserting the Lorentz scalar potential as $V_{1}(x)=-\hbar cA/|x|$ with a dimensionless coupling $A$ into Eq. (3.4) gives us
\begin{eqnarray}
\left\{-\frac{d^2}{dx^2}+Q^2(M^2-E^2)-\frac{2QMA}{|x|}+\frac{\xi^{(f,g)}}{x^2}\right\}\left[ \begin{array}{c} f(x)\\g(x)\end{array}\right]=0\,.
\end{eqnarray}
The upper indices $(f,g)$ of the constant $\xi$ indicates the upper component with $\xi^{f}=A(A-1)$, and the lower component with $\xi^{g}=A(A+1)$.

We obtain a second order equation for the upper component by changing the variable as $y=2\sqrt{Q^2(\mu^2-E^2)\,}|x|$
\begin{eqnarray}
\left\{\frac{d^2}{dy^2}-\frac{1}{4}+\frac{\mu A/\sqrt{\mu^2-E^2\,}}{y}-\frac{\xi^{(f,g)}}{y^2}\right\}f(y)=0\,,
\end{eqnarray}
The upper component can be taken of the form $e^{-y/2}y^{B}h(y)$ because of the asymptotic behaviours should be satisfied by a wave function. So, inserting this form into Eq. (3.6) gives for $h(y)$
\begin{eqnarray}
\left\{y\frac{d^2}{dy^2}+(2B-y)\frac{d}{dy}-\left(B-\frac{\mu A}{\sqrt{\mu^2-E^2\,}}\right)\right\}h(y)=0\,.
\end{eqnarray}
Eq. (3.7) is a confluent hypergeometric-type equation having the solutions [11]
\begin{eqnarray}
h(y) \sim \,_{1}F_{1}(B-\frac{\mu A}{\sqrt{\mu^2-E^2\,}};2B;y)
\end{eqnarray}
which can also be written in terms of the Laguerre polynomials $L_{n}^{k}(y)$ [14]. As stated in Ref. [15], by a detailed analyse on the constant $\xi^{(f,g)}$, we can give the bound states solutions, and the upper and lower components as
\begin{eqnarray}
f(y)=N_{1}y^{A}e^{-y/2}L_{n}^{2A-1}(y)\,,\nonumber\\
g(y)=N_{2}y^{A+1}e^{-y/2}L_{n-1}^{2A+1}(y)\,,
\end{eqnarray}
with
\begin{eqnarray}
E=\mp \mu\sqrt{1-\frac{A^2}{(n+A)^2}\,}\,\,;n=0, 1, 2, \ldots
\end{eqnarray}

We are now ready to write the eigenvalues for the case where the potential field is strong which means $A \gg 1$ going to further for getting an analytical expression for the partition function below. For this case, we have $E \sim \mp \mu\sqrt{\frac{2n}{A}\,}$, and if the potential field becomes extremely strong the eigenvalues go to zero [15]. In the next section, we will use the above result to write the partition function for particle eigenvalues.

\subsection{T\lowercase{hermal} Q\lowercase{uantities}}
The partition function of the system can be written as a summation over all the quantum states [1]
\begin{eqnarray}
Z(\beta)=\sum_{n=0}^{\infty}e^{-\frac{1}{\bar{\mu}}\,E_{n}}\,,
\end{eqnarray}
with $\bar{\mu}=1/\beta \mu$, and $\beta=1/k_{B}T$ where $k_{B}$ Boltzmann constant, and $T$ is temperature in Kelvin.

In order to evaluate the partition function analytically, we can use the following integral equality over the contour $C$ [6, 7, 9]
\begin{eqnarray}
e^{-y}=\frac{1}{2\pi i}\int_{C}y^{-t}\Gamma(t)dt\,,
\end{eqnarray}
where $\Gamma(z)$ is the Euler Gamma function [11]. With the help of Eq. (3.12), and by using the Riemann Zeta function defined as $\zeta(t)=\sum_{n=0}^{\infty}1/n^{t}$ in Eq. (3.11) [11], we write
\begin{eqnarray}
\sum_{n=0}^{\infty}e^{-\frac{1}{\bar{\mu}}\,\sqrt{2an\,}}=\frac{1}{2\pi i}\int_{C}(\bar{\mu})^{t}(2a)^{-t/2}\zeta(t/2)\Gamma(t)dt\,,
\end{eqnarray}
with $a=1/A$. We see that Eq. (3.13) has two simple poles at $t=2$, and $t=0$. By applying the residue theorem, the wanted partition function is obtained as
\begin{eqnarray}
Z(\bar{\mu})=\frac{\bar{\mu}^2}{2a}+\frac{1}{2}\,.
\end{eqnarray}

Before going further, from this result, we tend to write here the reduced thermal functions explicitly
\begin{eqnarray}
\bar{F}(\bar{\mu})&=&\frac{F}{\bar{\mu}}=-\bar{\mu}\text{ln}\left(\frac{\bar{\mu}^{2}+a}{2a}\right)\,,\nonumber\\
\bar{U}(\bar{\mu})&=&\frac{U}{\mu}
=\frac{2\bar{\mu}^{3}}{\bar{\mu}^{2}+a}\,,\nonumber\\
\bar{C}(\bar{\mu})&=&\frac{C}{k_{B}}=\frac{2\bar{\mu}^{2}(\bar{\mu}^{2}+3a)}{(\bar{\mu}^{2}+a)^{2}}\,,
\end{eqnarray}

Let us first analyze the case of high temperatures corresponding to $\beta \ll 1$. Eq. (3.14) gives the following results for high temperatures
\begin{eqnarray}
Z \sim \bar{\mu}^{2}/2a\,\,;U \sim 4\bar{\mu}\,\,;C \sim 2\,.
\end{eqnarray}
We observe that the partition function still depends also to the potential parameter $a$ while the dependency of the mean energy, and the specific heat disappear for high temperatures, which can be seen in Figs. (6)-(8) below.

We observe that the thermodynamic quantities in Eq. (3.15) depend also the coupling parameter $a$. So, we give our all numerical results as the variation of them versus the temperature for three different values of potential parameter, namely, $a=1$, $a=0.5$ and $a=0.1$, in Figs. (6)-(8). Fig. (6) shows that the Helmholtz free energy increase with increasingly value of $a$. In Fig. (7), we see that the effect of the coupling parameter on the mean energy is more apparent for nearly low temperatures. On the other hand, the plots for different $a$-values for the mean energy are closing to each other which means that the dependence of the mean energy on $a$ disappears for high temperatures. We give the variation of the specific heat according to the temperature in Fig. (8) where it has an upper value while the temperature increases.

\section{C\lowercase{onclusions}}

We have studied the thermal functions for the Klein-Gordon equation with a Lorentz scalar potential having a linear form, and for the Dirac equation with a Lorentz scalar, inverse-linear potential for the case where the potential field is strong. For this aim, we have considered the case where the positive part of the energy spectrum gives a contribution to the partition function. All important thermodynamics quantities have been computed by a method based on the Euler-MacLaurin formula for the KG solutions, and a method based on Euler Gamma and the Riemann Zeta functions for the Dirac solutions, respectively. The results for high temperatures are also given for the specific heat, and the mean energy.

\section{A\lowercase{cknowledgments}}
One of authors (A.A.) thanks Prof Dr Andreas Fring from City University London and the Department of Mathematics for hospitality. This research was partially supported by the Scientific and Technical Research Council of Turkey and through a fund provided by University of Hacettepe.

\newpage

\begin{figure}
\centering
\includegraphics[height=3.5in, width=5in, angle=0]{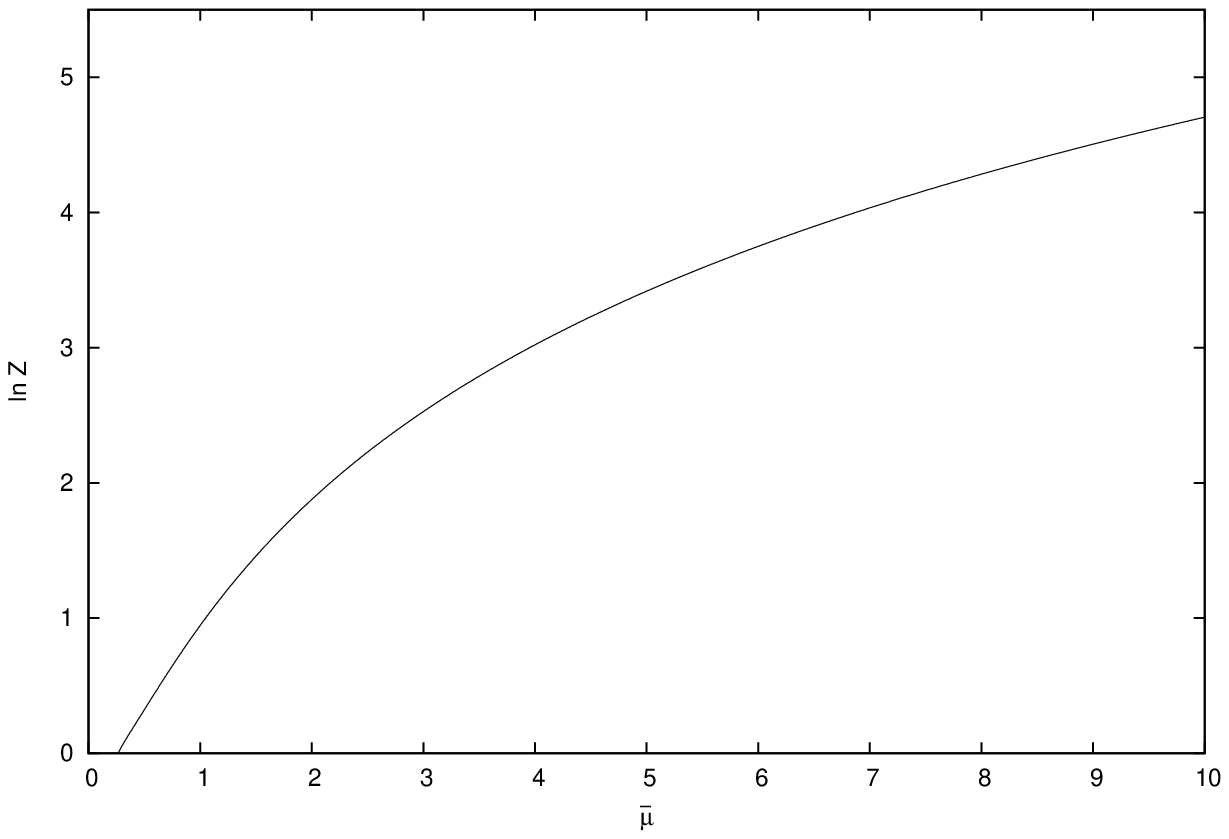}
\caption{The variation of $\text{ln}Z$ versus $\bar{\mu}$.}
\end{figure}

\begin{figure}
\centering
\includegraphics[height=3.5in, width=5in, angle=0]{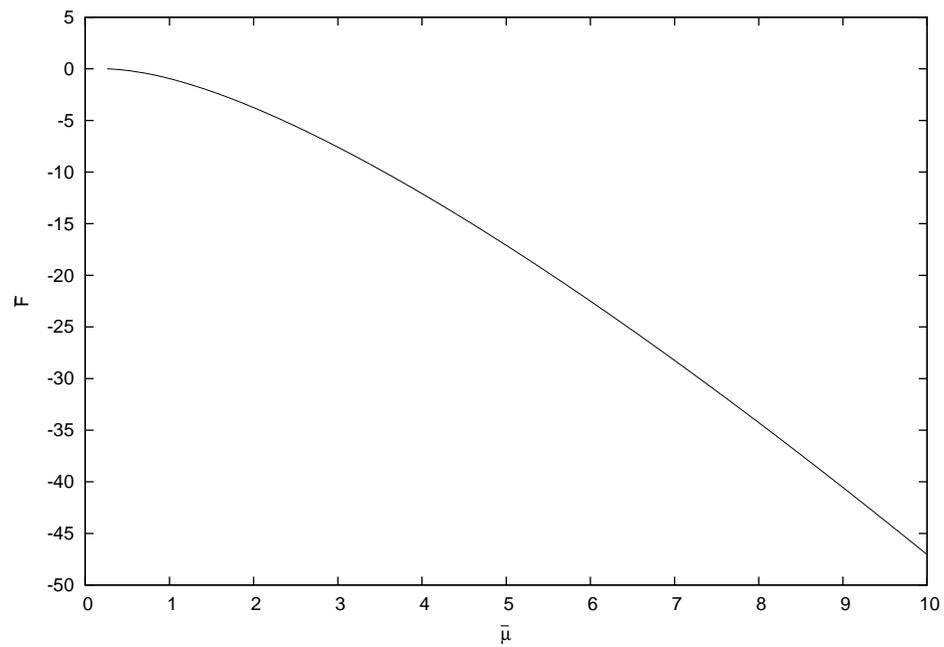}
\caption{The free energy for linear potential versus $\bar{\mu}$.}
\end{figure}

\newpage

\begin{figure}
\centering
\includegraphics[height=3.5in, width=5in, angle=0]{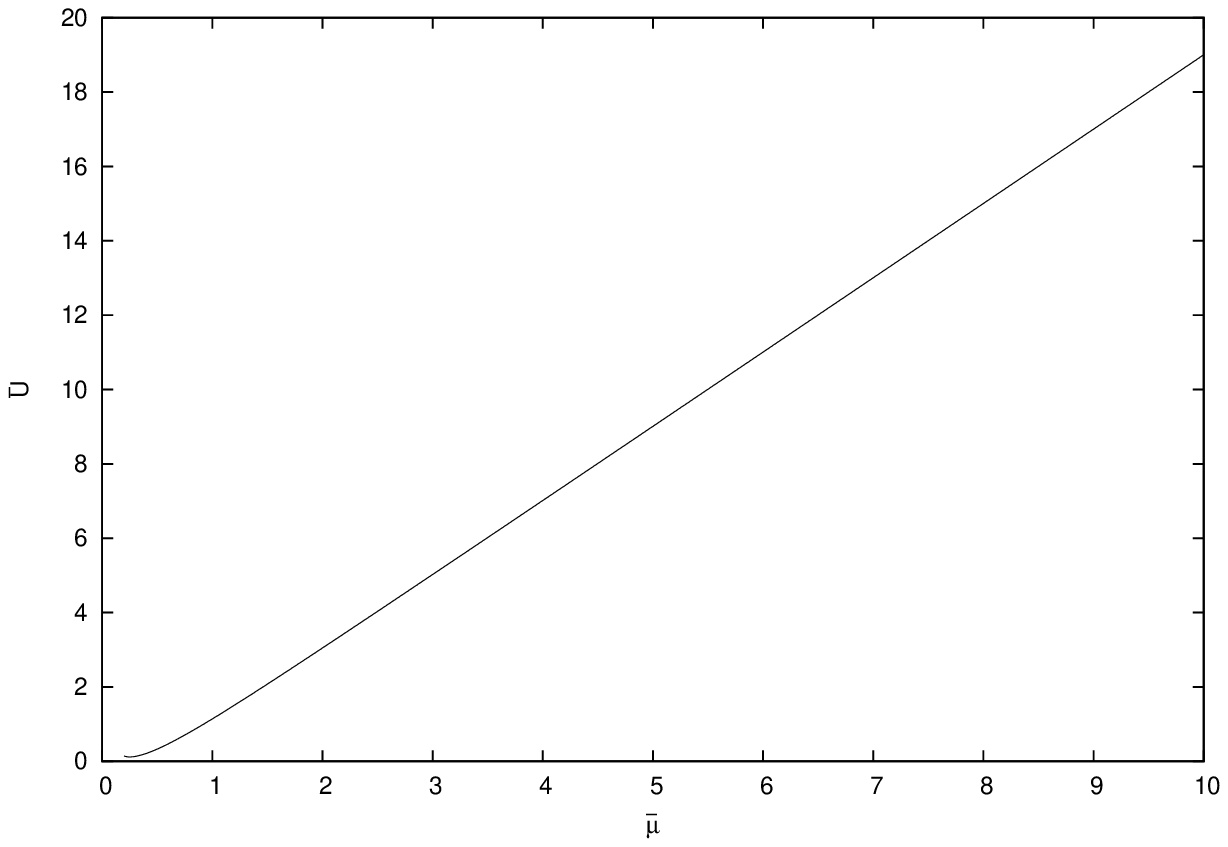}
\caption{The mean energy for linear potential versus $\bar{\mu}$.}
\end{figure}

\begin{figure}
\centering
\includegraphics[height=3.5in, width=5in, angle=0]{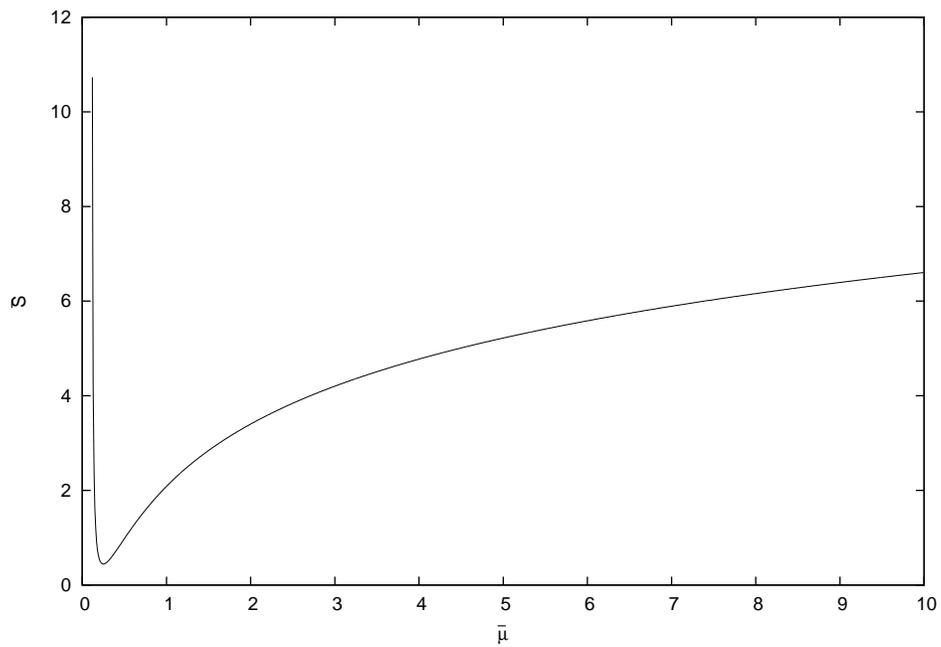}
\caption{The entropy for linear potential versus $\bar{\mu}$.}
\end{figure}

\newpage

\begin{figure}
\centering
\includegraphics[height=3.5in, width=5in, angle=0]{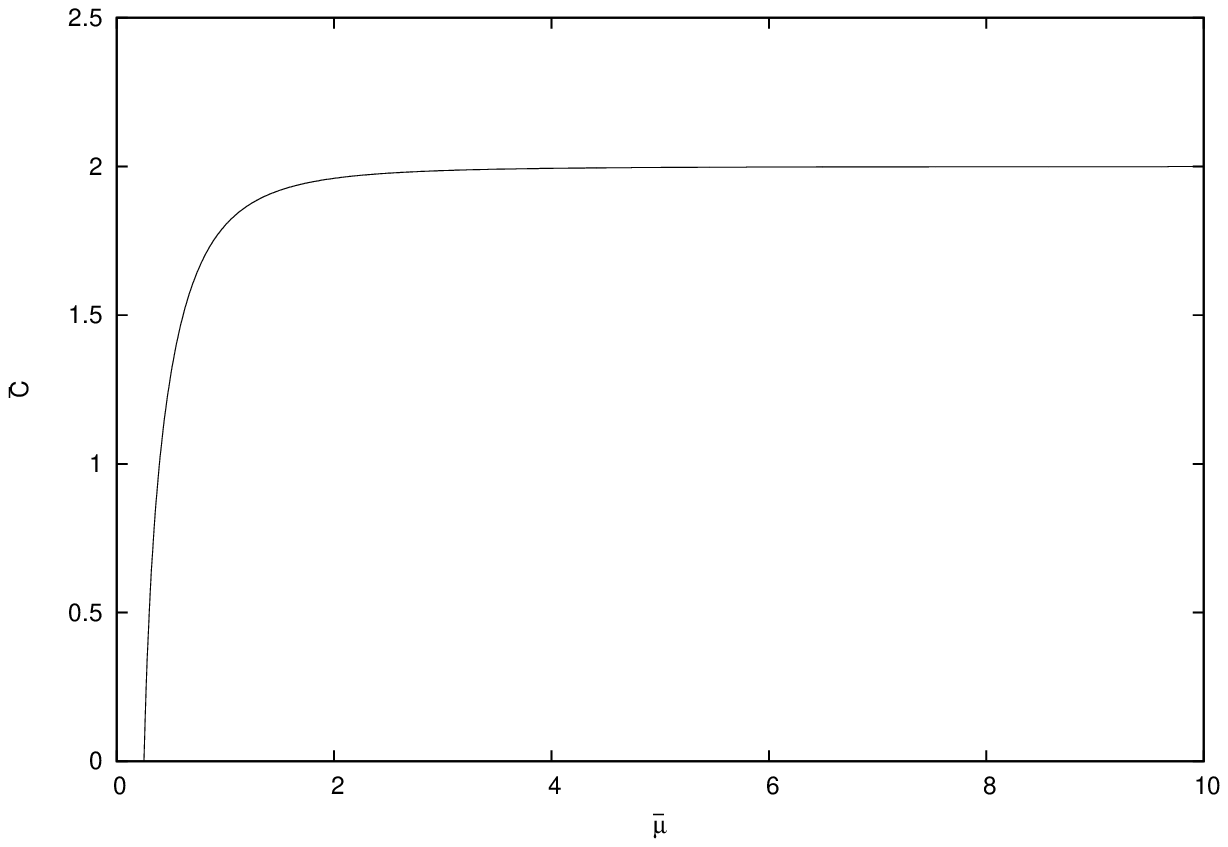}
\caption{The specific heat for linear potential versus $\bar{\mu}$.}
\end{figure}

\newpage

\begin{figure}
\centering
\includegraphics[height=3.5in, width=5in, angle=0]{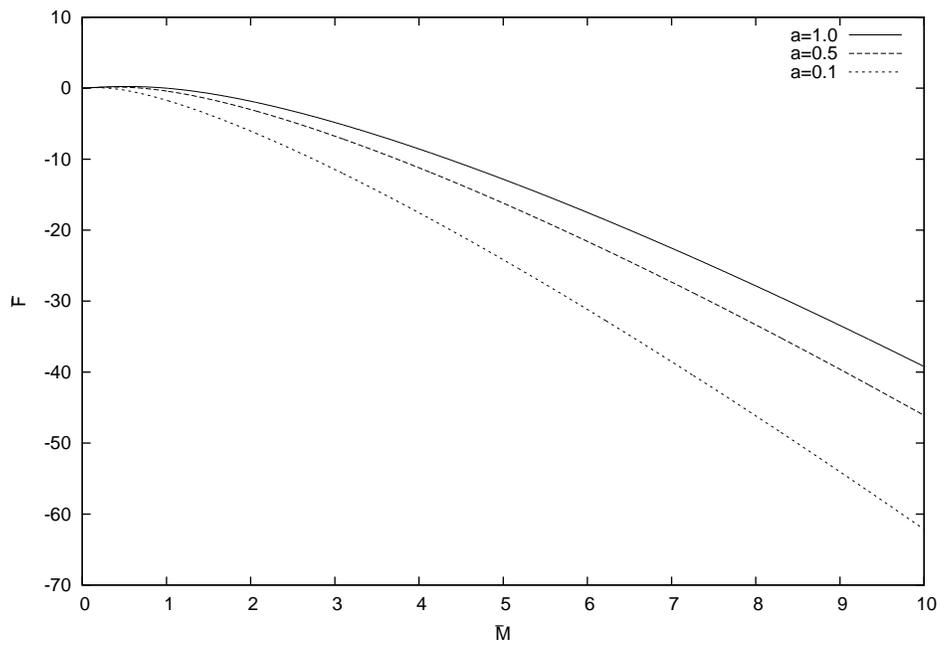}
\caption{The Helmholtz free energy for the inverse-linear potential versus $\bar{\mu}$.}
\end{figure}

\begin{figure}
\centering
\includegraphics[height=3.5in, width=5in, angle=0]{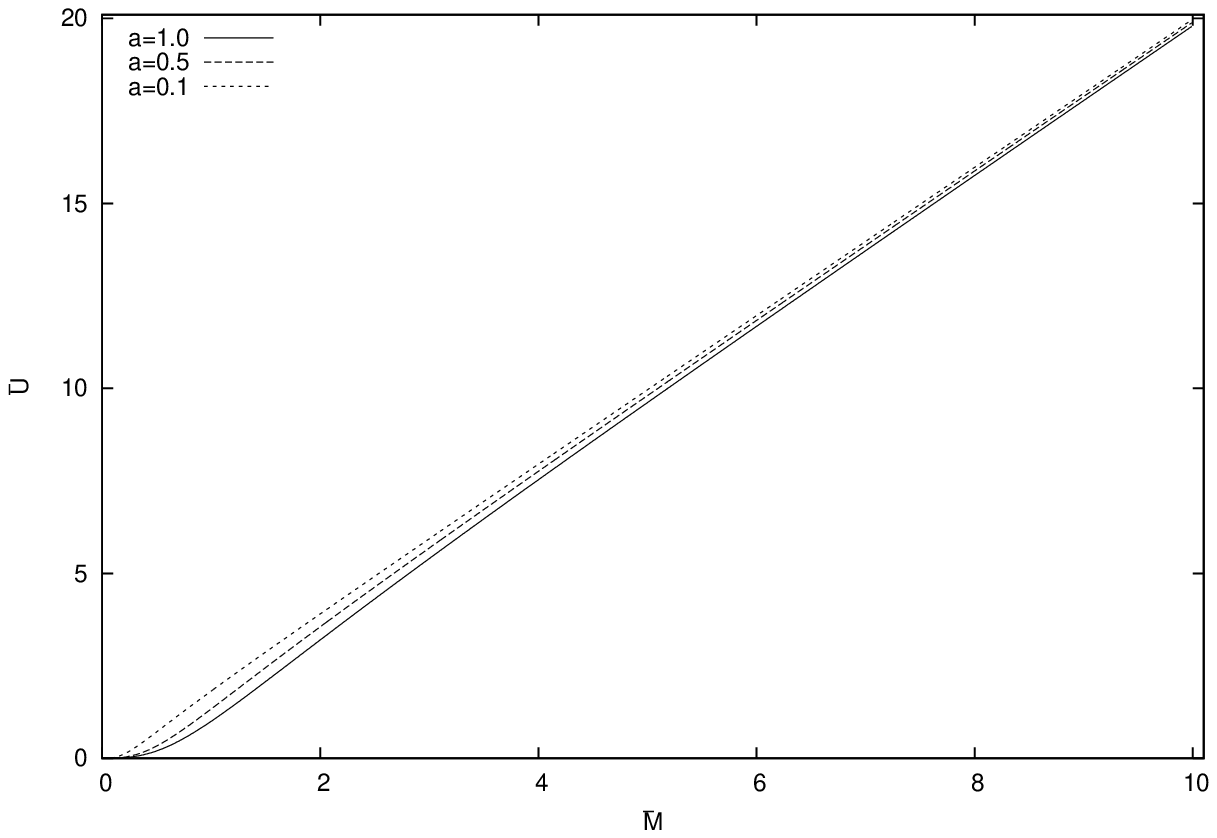}
\caption{The mean energy for the inverse-linear potential versus $\bar{\mu}$.}
\end{figure}

\newpage

\begin{figure}
\centering
\includegraphics[height=3.5in, width=5in, angle=0]{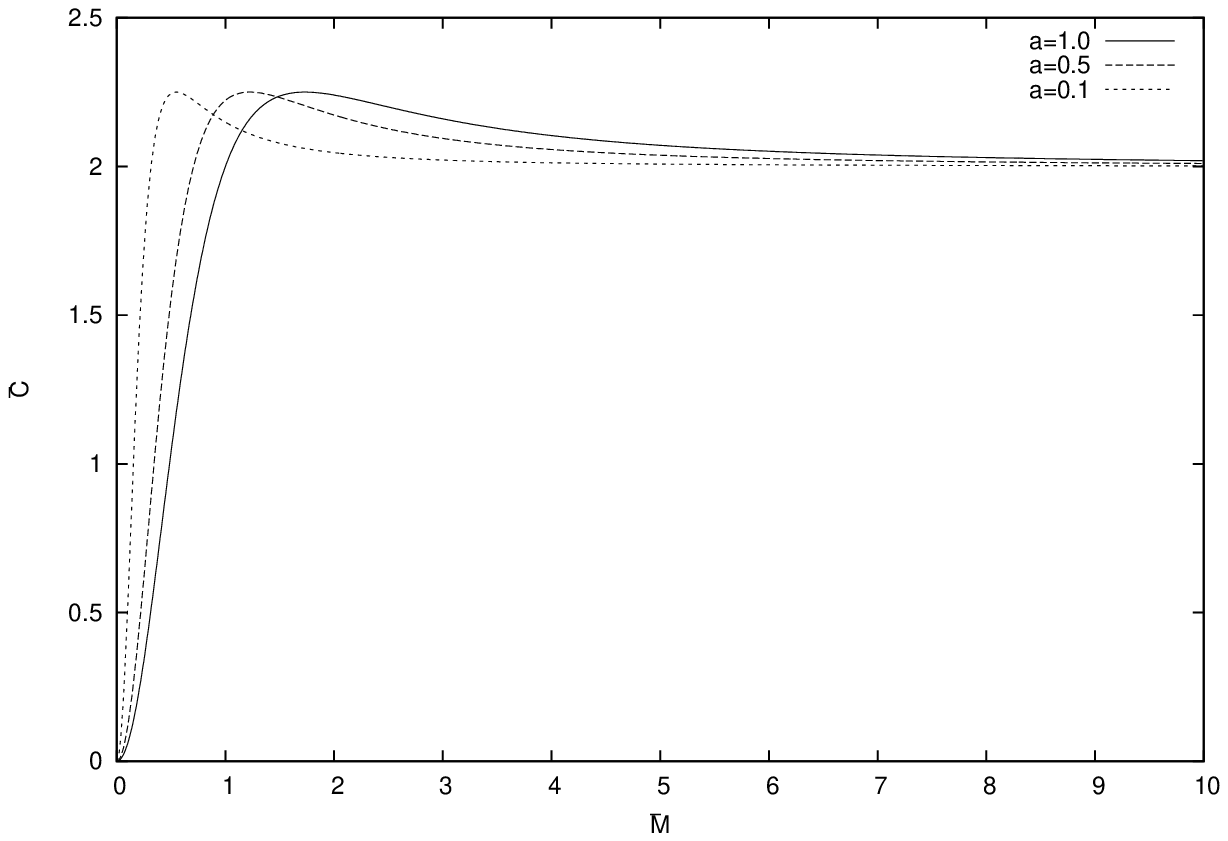}
\caption{The specific heat for linear potential versus $\bar{\mu}$.}
\end{figure}

\end{document}